\documentclass{aa}
\usepackage{natbib}
\usepackage{txfonts}
\usepackage{graphics}
\usepackage{amssymb}
\begin{document}
\title{The effect of a strong external radiation field on protostellar envelopes in Orion}
\author{J.~K. J{\o}rgensen\inst{1,2}\thanks{Current address: Harvard-Smithsonian Center for Astrophysics} \and D. Johnstone\inst{3,4} \and E.~F. van Dishoeck\inst{1} \and S.~D. Doty\inst{5}} 

\institute{Leiden Observatory, P.O. Box 9513, NL-2300 RA Leiden, The Netherlands \and Harvard-Smithsonian Center for Astrophysics, 60 Garden Street MS42, Cambridge, MA 02138, USA \and
National Research Council Canada, Herzberg Institute of Astrophysics, 5071 West Saanich Rd, Victoria, BC, V9E 2E7, Canada \and
Department of Physics \& Astronomy, University of Victoria, Victoria, BC, V8P 1A1, Canada \and 
Department of Physics and Astronomy, Denison University, Granville, OH 43023, USA}

\offprints{Jes K.\,J{\o}rgensen} \mail{jjorgensen@cfa.harvard.edu}
\date{Received 8 March 2005 / Accepted 5 December 2005}

\abstract{We discuss the effects of an enhanced interstellar radiation
  field (ISRF) on the observables of protostellar cores in the Orion
  cloud region. Dust radiative transfer is used to constrain the
  envelope physical structure by reproducing SCUBA 850~$\mu$m
  emission. Previously reported $^{13}$CO, C$^{17}$O and H$_2$CO line
  observations are reproduced through detailed Monte Carlo line
  radiative transfer models. It is found that the $^{13}$CO line
  emission is marginally optically thick and sensitive to the physical
  conditions in the outer envelope. An increased temperature in this
  region is needed in order to reproduce the $^{13}$CO line strengths
  and it is suggested to be caused by a strong heating from the
  exterior, corresponding to an ISRF in Orion $10^3$ times stronger
  than the ``standard'' ISRF. The typical temperatures in the outer
  envelope are higher than the desorption temperature for CO. The
  C$^{17}$O emission is less sensitive to this increased temperature
  but rather traces the bulk envelope material. The data are only fit
  by a model where CO is depleted, except in the inner and outermost
  regions where the temperature increases above 30-40~K. The fact that
  the temperatures do not drop below $\approx 25$~K in any of the
  envelopes whereas a significant fraction of CO is frozen-out suggest
  that the interstellar radiation field has changed through the
  evolution of the cores. The H$_2$CO lines are successfully
  reproduced in the model of an increased ISRF with constant
  abundances of 3--5$\times 10^{-10}$.

}
\maketitle

\section{Introduction}
Giant molecular clouds are the formation sites of massive stars in our
Galaxy with the nearby Orion molecular clouds being the prime
candidates for detailed studies of the earliest protostellar stages.
An interesting difference for the studies of these cores compared to
isolated systems is the large number of OB stars in the immediate
vicinity: the UV radiation from these early-type stars ionizes their
surrounding material within a few parsecs, and also affects the
thermal balance and chemistry in intermediate and low-mass
protostellar cores distributed over much larger scales. As recent
studies indicate that large numbers of solar-type stars may be formed
in these regions, it is of great importance to address the feedback
between high- and low-mass star formation, in particular, for
comparison with low-mass protostars formed in relative isolation in
clouds such as Taurus.

In this paper we present a radiative transfer study of three
intermediate mass protostars in Orion from the sample of
\cite{johnstone03}. The physical properties of their envelopes are
established from 1D dust radiative transfer modeling, and CO isotopic
and H$_2$CO line observations are analyzed using Monte Carlo line
radiative transfer. In particular, we discuss the importance of
heating the protostellar envelope via an enhanced external radiation
field and the constraints on this heating from optically thick CO
isotopomers.

\cite{li03} recently estimated the gas kinetic temperatures for a
sample of pre-stellar cores in Orion using inversion transitions of
NH$_3$. They found that these cores had lower temperatures than their
surroundings, which they attributed to the impact of the strong
interstellar radiation field (ISRF) in the Orion region. This is
reminiscent of the situation for pre-stellar cores in more quiescent
star forming regions where temperature gradients due to a standard
interstellar radiation field are found in radiative transfer models
reproducing submillimeter continuum maps \citep[e.g.,][]{evans01}.
\cite{wilson99orion} suggested that the ratios of peak temperatures of
optically thick lines with different critical densities could be used
as a diagnostic of temperatures. Applied to cores in Orion their
results also suggested strong temperature gradients in the region
around Orion BN/KL with more dense gas traced by NH$_3$ inversion
lines colder ($\sim 24$~K) than less dense gas ($\sim 40$~K) probed by
CO lines. In contrast \citeauthor{wilson99orion} found similar
temperatures probed by the NH$_3$ and CO lines for cores further south
from the Orion BN/KL region and they argued that these cores are
dominated by internal rather than external heating. This illustrates
that an important point for the thermal balance of protostellar
objects is whether external heating can compete with the central
source luminosity.

The strong interstellar radiation field may also be reflected in the
chemistry, for example by increasing the photodissociation and
photoionization. The strength of the UV field can for example be
probed by the emission of C$^+$ and has been applied to the Orion
clouds where UV fields enhanced by factors of $10^3-10^5$ have been
suggested
\citep[e.g.,][]{tielens85b,tielens85a,burton90,stacey93,moorkerjea03}. The
strong UV field will also lead to enhanced abundances of electrons and
thus abundance decreases of molecular ions, such as HCO$^+$ and
N$_2$H$^+$ that otherwise work as destroyers of common
molecules. Support for this is found within the OMC1 cloud core
located immediately behind the Orion Nebula Cluster where
\cite{ungerechts97} infer abundance gradients along the cloud, with
increasing abundances of species such as HCN, CH$_3$OH, HC$_3$N, and
SO toward the Orion BN/KL region.

The astrochemistry study by \cite{johnstone03} considered a selection
of cornerstone molecules in order to quantify the range of conditions
for which individual molecular line tracers provide physical or
chemical information. The morphological study compared a variety of
locations along the Integral Shaped Filament (ISF) in Orion A
\citep{bally87,johnstone99}, chosen to represent a range of physical
conditions including enshrouded protostars, a bright PDR knot, and a
shock front. The main conclusion of the paper was that the abundances
of many molecular species correlate with source energetics, likely a
result of the importance of temperature dependent desorption in
maintaining gas-phase molecules. A significant finding of the study
was the need for a warm outer envelope for all of the protostellar
sources, as expected given the proximity of numerous O and B
stars. The study suffered, however, from a simplistic treatment of the
molecular line abundances derived from single temperature and density
statistical equilibrium calculations and thus was unable to fully
consider the effect of an enhanced radiation field on the outer
regions of the embedded protostars.

This paper presents a continuation of the \cite{johnstone03} study,
performing a detailed radiative transfer analysis of the physical and
chemical properties of a subsample of the regions studied in that
paper. We concentrate on the embedded intermediate mass protostellar
sources, MMS6, MMS9, and FIR4 where the assumption of constant density
and temperature throughout the envelope is most suspect. All three
sources have large envelope masses, $M_{\rm env} > 10 M_\odot$.  The
implied dust temperatures, $T_{\rm d}$, for the sources are $>15$~K
while the $^{13}$CO lines have peak brightness temperatures
$\gtrapprox 30\,$K which is difficult to produce {\it unless} the
cores are bathed in an enhanced interstellar radiation field. The
properties of the cores studied in this paper are summarized in
Table~\ref{sample} with further details presented in
\cite{johnstone03}. Even though the focus of this paper is on a few
selected cores in Orion, a lot of the results brought up in this
discussion are valid beyond these specific sources.

This paper is laid out as follows: sect.~\ref{model} introduces the
general problem, including the observational indications for a strong
interstellar radiation field, and describes the modeling approach,
including the possible constraints from continuum and line
observations. Sect.~\ref{discuss} then discusses possible refinements
and implications of the models, in particular, the resulting CO
abundance structures and the reproduction of H$_2$CO multi-transition
observations. The evolutionary implications are discussed in
Sect.~\ref{s:implic}.  Sect.~\ref{s:conclusion} concludes the paper
suggesting possible further tests and future work. 

\begin{table*}
\caption{Sample of sources in Orion. For details see
  \cite{johnstone03}. }\label{sample}
\begin{center}
\begin{tabular}{lllllll}\hline\hline
Source       &                                 & MMS6 & MMS9 & FIR4  \\ \hline
$L_{\rm bol}$ &[$L_\odot$]                     & 60   & 90   & 400       \\
Projected distance to the Trapezium$^a$  &    [pc] & 2.9  & 2.3    & 1.7    \\
$S_{850 \mu{\rm m}}$ (peak) & [Jy beam$^{-1}$] & 7.5  & 2.5  & 7.5      \\
$I$($^{13}$CO 3--2)$^b$ &[K~km~s$^{-1}$]              & 40.7 & 59.1 & 76.2 \\
$I$(C$^{17}$O 3--2)$^b$ & [K~km~s$^{-1}$]             & 4.78 & 2.88 & 3.97 \\
$I$(H$_2$CO $3_{03}-2_{02}$)$^b$ & [K~km~s$^{-1}$]    & 5.33 & 3.07 & 9.82  \\
$I$(H$_2$CO $3_{22}-2_{21}$)$^b$ & [K~km~s$^{-1}$]    & 0.53 & 0.28 & 2.96  \\
$I$(H$_2$CO $5_{05}-4_{04}$)$^b$ & [K~km~s$^{-1}$]    & 2.56 & 0.94 & 5.24  \\ \hline
\end{tabular}
\end{center}

$^a$Assuming a distance to Orion of 450~pc. $^b$Line intensity,
$I=\int T_{\rm MB}\,{\rm d}v$.
\end{table*}

\section{Models: constraining the ISRF}\label{model}
Since the protostellar cores in Orion appear warmer than expected from
heating by only the low luminosity central sources, they require the
enhanced interstellar radiation field produced by the nearby OB stars
for additional heating. In this section we present continuum and line
radiative transfer models which describe the thermal structure of
these cores and can be used for comparison to the observations. First,
however, we describe direct evidence of the contribution from external
heating of these cores - in particular from $^{13}$CO isotopic lines and
dust continuum observations obtained with the James Clerk Maxwell
Telescope in a $\approx$~15\arcsec\ beam \citep{johnstone03,paperii}.

\subsection{$^{13}$CO 3--2 and ISRF}
\begin{figure}
\resizebox{\hsize}{!}{\includegraphics{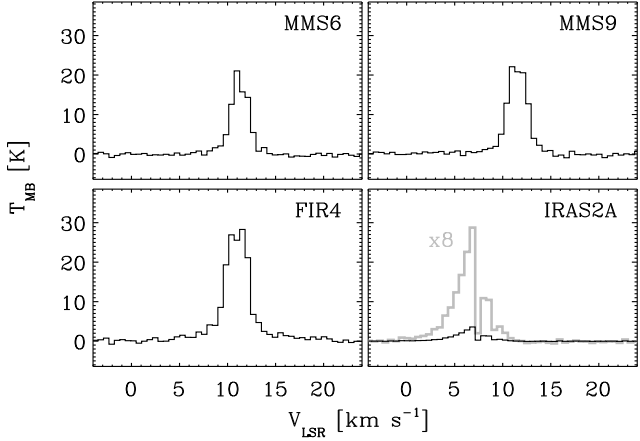}}
\caption{JCMT $^{13}$CO 3--2 spectra from the three sources in
    Orion compared to NGC~1333-IRAS~2A.}\label{s:tcospectra}
\end{figure}

Fig.~\ref{s:tcospectra} compares the $^{13}$CO 3--2 spectra toward the
three sources.  Also shown is the low-mass protostar, NGC~1333-IRAS~2A
\citep{paperii} which has a lower luminosity (16~$L_\odot$) than the
Orion sources but also is located at shorter distances (220 pc). The
peak temperatures for the three Orion sources are significantly higher
than what is observed for NGC~1333-IRAS~2A. Also note that the lines
are relatively Gaussian with less significant self-absorption. Since
$^{13}$CO 3--2 is optically thick for the column densities representative
of the studied cores, these spectra indicate that a relatively warm
foreground layer must be present to prevent any self-absorption dip
from the larger scale cold envelope. Comparing the $^{13}$CO and
C$^{17}$O 3--2 spectra from \cite{johnstone03} provides further evidence:
the $^{13}$CO lines are about a factor 2 broader than those of C$^{17}$O,
suggesting that the lines are optically thick.

If the $^{13}$CO 3--2 line is optically thick, the peak temperature
provides a measures of the excitation temperature in the region of the
envelope where the optical depth reaches about unity. For the high
densities in these envelopes the CO gas is in LTE
\citep[e.g.,][]{jorgensen02} and the excitation temperature directly
measures the gas kinetic temperature. Furthermore at these densities
the gas and dust temperatures are expected to be closely coupled
\citep[e.g.,][]{ceccarelli96,doty97} and can as a first approximation
be assumed identical. In the following section we carry out the
detailed line radiative transfer to calculate the full non-LTE line
excitation and explore the relevance of the dust/gas temperature
decoupling.

Correcting the peak brightness temperatures $T_{\rm peak}$=17--28~K
\citep{johnstone03} for the Rayleigh-Jeans approximation, indicates
gas temperatures of $T_{\rm g}$=24--35~K, assuming that the emission
fills the beam. Such temperatures are, however, impossible to
reconcile with the central star being the sole source of
luminosity. For a spherically symmetric, optically thin envelope
heated by a single central source of luminosity, the dust temperature
distribution is approximated by:
\begin{equation}
T_{\rm d}(r)=60\left(\frac{r}{2\times10^{17}~{\rm cm}}\right)^{-q}\left(\frac{L_{\rm bol}}{10^5~L_\odot}\right)^{q/2}
\label{e:chandlertemp}
\end{equation}
where $q$ is related to the power-law index $\beta$ of the dust
opacity law, $\kappa=\kappa_0(\nu/\nu_0)^\beta$, through
$q=2/(4+\beta)$, \citep{scoville76,doty94,chandler98,chandler00}.
This would be the temperature probed by the $^{13}$CO 3--2 lines if the
emission comes from a shell with diameter larger than the JCMT
beam which is very optically thick in these lines. To reach a specific
temperature $T_{\rm d}$ at this radius the required luminosity is
determined from Eq.~(\ref{e:chandlertemp}):
\begin{equation}
L_{\rm bol}=6\times 10^5\,\left(\frac{T_{\rm d}}{60~{\rm K}}\right)^{2/q}\left(\frac{d}{460~{\rm pc}}\right)^2\,  L_\odot
\label{e:lumargument}
\end{equation}
Therefore, to produce peak temperatures $T_{\rm peak}$=17--28~K in the
JCMT beam at 330~GHz at the distance of Orion (460~pc), central
sources of $L\sim 6\times 10^3-4\times 10^4~L_\odot$ are needed,
significantly higher than the luminosity of the observed sources. In
contrast, the $T_{\rm peak}\approx 4$~K for NGC~1333-IRAS~2A requires
only 15-20~$L_\odot$, comparable to the bolometric luminosity of the
source \citep{jorgensen02}. It therefore appears that the only way to
explain the $^{13}$CO 3--2 line emission for the Orion sources is to
introduce a strong external source of heating.

\subsection{Continuum models}
\begin{figure*}
\resizebox{\hsize}{!}{\includegraphics{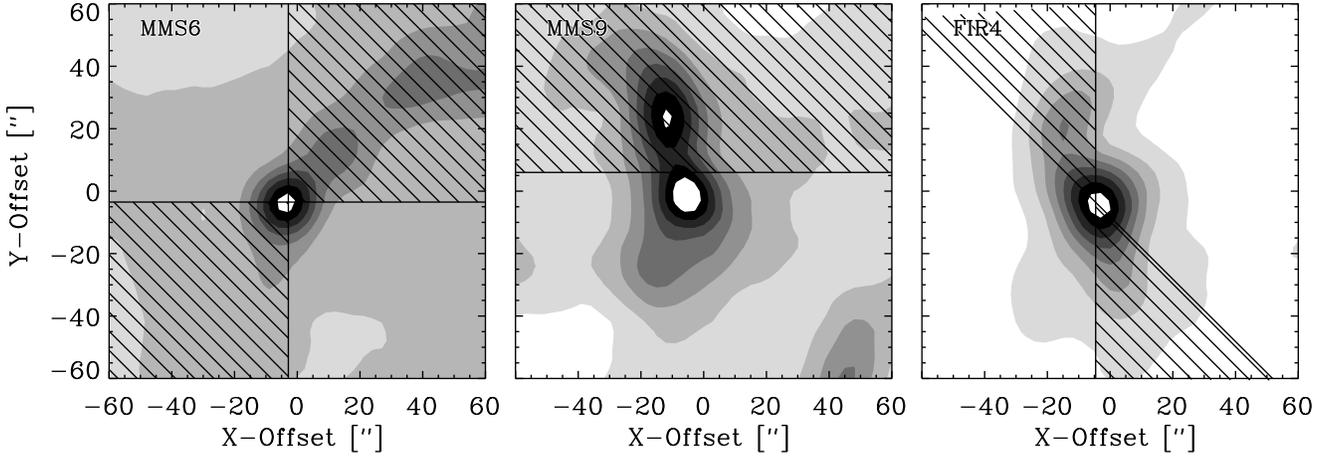}}
\caption{SCUBA 850~$\mu$m maps of the three Orion sources (from left
to right: MMS6, MMS9 and FIR4). The regions which have been excluded
when extracting the brightness profiles are hatched
out.}\label{extract_prof}
\end{figure*}

To establish the physical structure of each core we adopted the same
approach as in \cite{jorgensen02}, using the 1D radiative transfer
code Dusty \citep{dusty} to simulate 850~$\mu$m SCUBA images and to
constrain the density distribution of each core by comparing to
observations. As in \citeauthor{jorgensen02} the opacities for
coagulated dust grains with thin ice mantles at a density of
$10^6$~cm$^{-3}$ from \cite{ossenkopf94} were adopted. The density
profile is assumed to have a single radial power-law, $n=n_0
(r/r_0)^{-p}$. The power-law index, $p$, is constrained through
comparison between observed and modeled brightness profiles. The Orion
protostars are located in the dense ridge of the ISF
\citep{johnstone99} complicating the extraction of a unique brightness
profile. Greyscale images of each protostar are shown in
Figure~\ref{extract_prof}. The hashed regions in each plot were not
included in the circular averaging procedure used to define the
brightness profiles. The flux at 850~$\mu$m for each source from
\cite{johnstone03} was used to constrain the overall dust content in
the core, that is the optical depth at the fiducial wavelength
\citep[see, e.g., discussion in][]{jorgensen02}. Following
\cite{evans01} we adopt the average ISRF from the solar neighborhood
of \cite{black94} with the UV field from \cite{draine78} added and
scale this at every wavelength with a constant factor, $\chi_{\rm
  ISRF}$ (i.e., $\chi_{\rm ISRF}=1$ for the standard ISRF).

\begin{figure}
\resizebox{\hsize}{!}{\includegraphics{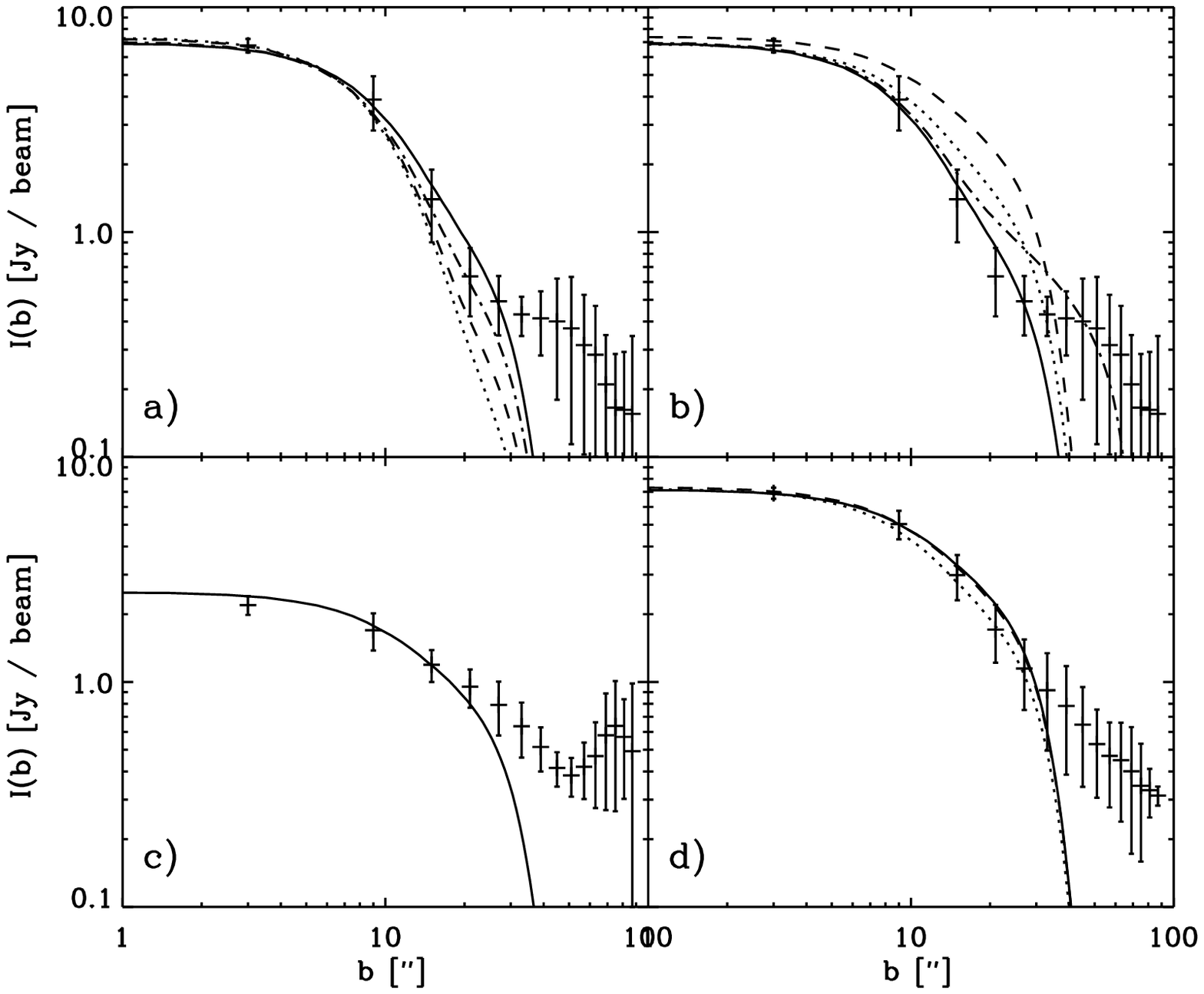}}
\caption{Fits to the images of each of the Orion cores. a) MMS6 models
  with varying ISRF ($\chi_{\rm ISRF}$ of 1, 10, 100 and 1000 with
  dotted, dashed, dotted-dashed and solid lines, respectively). b)
  MMS6 models with varying density slope ($p=1.5, 1.75, 2.0$ with
  dashed, dotted and solid lines, respectively). The
    dashed-dotted line indicates a model with a density slope of 2.0
    and an outer radius of 30,000 AU rather than 15,000 AU. c) MMS9
  model. d) FIR4 model with varying ISRF ($\chi_{\rm ISRF}$=1000 and
  10000 with dotted and solid lines,
  respectively).}\label{f:dustprofile}
\end{figure}

Fig.~\ref{f:dustprofile} illustrates the differences in the brightness
profiles produced by varying the density profile slopes and the impact
of the external interstellar radiation field. As shown, when including
the ISRF a steeper density profile is required to fit the data. The
best fit for MMS6 is a steep $r^{-2}$-envelope, whereas MMS9 and FIR4
have flatter $r^{-1.5}$-envelopes.
\begin{table*}
\caption{Summary of the tested models.}
\label{sum2}
\begin{tabular}{lllllllll} \hline\hline 
        & ISRF   & $p$  & $\tau_{100}$ & $n_i$   & $\langle T \rangle$$^{a}$ & $I(^{13}$CO$)$$^{b}$ & $I($C$^{17}$O$)$$^{b}$ & In Figures \\ 
        &        &      &         & 10$^8$~cm$^{-3}$ & K & K~km~s$^{-1}$           & K~km~s$^{-1}$           &  \\ \hline
MMS6    & 1      &  2.0   &  8.0  & 37  & 19 & 22.3 (56)   & 10.2 (2.1)   & \ref{f:dustprofile}a: dotted \\
        & 10     &  2.0   &  7.2  & 33  & 21 & 27.0 (51)   & 9.95 (1.9)   & \ref{f:dustprofile}a: dashed \\
        & 100    &  2.0   &  6.5  & 30  & 26 & 37.4 (46)   & 10.6 (1.7)   & \ref{f:dustprofile}a: dotted-dashed \\
        & 1000   &  1.5   &  2.4  & 5.9 & 31 & 58.6 (37)   & 17.4 (1.4)   & \ref{f:dustprofile}b: dashed \\
        & 1000-D &  1.5   &  2.4  & 5.9 & 31 & 55.4 (37)   & 17.2 (1.4)   & \\
        & 1000   &  1.75  &  3.9  & 13  & 32 & 56.2 (39)   & 14.2 (1.5)   & \ref{f:dustprofile}b: dotted \\
        & 1000*  &  2.0   &  5.6  & 26  & 36 & 53.6 (40)   & 11.1 (1.5)   & \ref{f:dustprofile}a,b: solid \\
        & 1000-2 &  2.0   &  5.6  & 22  & 37 & 50.6 (40)   & 11.4 (1.5)   & \ref{f:dustprofile}b: dotted-dashed \\
        & 10000  &  2.0   &  4.5  & 21  & 54 & 73.4 (28)   & 10.2 (1.1)   & \\
MMS9    & 1      &  1.5   &  1.1  & 2.7 & 17 & 34.6 (12)   & 8.30 (0.47)  & \\
        & 1000*  &  1.5   &  0.7  & 1.7 & 36 & 64.3 (6.7)  & 7.13 (0.26)  & \ref{f:dustprofile}c: solid \\
        & 10000  &  1.5   &  0.45 & 1.1 & 57 & 70.9 (3.5)  & 4.92 (0.13)  & \\
FIR4    & 1      &  1.5   &  2.3  & 5.6 & 22 & 50.3 (16)   & 17.5 (0.61)  & \\
        & 1000   &  1.5   &  1.8  & 4.4 & 34 & 81.3 (11)   & 16.4 (0.43)  & \ref{f:dustprofile}d: dotted \\
        & 5000   &  1.5   &  1.5  & 3.7 & 45 & 99.2 (8.5)  & 14.7 (0.32)  & \ref{f:dustprofile}d: solid \\
        & 10000* &  1.5   &  1.3  & 3.2 & 51 & 105.5 (6.9) & 12.9 (0.26)  & \\ \hline
\end{tabular}

Notes: MMS6 model marked with ``-D'' is with gas and dust temperature
decoupling taken into account. The inner and outer radii are 50
  and 15,000~AU, respectively except for the MMS6 model marked with
  ``-2'' which has an outer radius of 30,000~AU. Models marked with
``*'' are the best fit models used in Table~\ref{sum3}. Models for
MMS6 have a constant turbulent broadening of 0.7~km~s$^{-1}$, those for MMS9
and FIR4 1.0~km~s$^{-1}$. $^{a}$Mass-weighted temperature: $\langle T
\rangle=\int T(r)\,n(r)\,4\pi r^2\, {\rm d}r/\int n(r)\,4\pi r^2\,
             {\rm d}r$. $^{b}$Modeled $J=3-2$ line intensity in 15$''$
             JCMT beam with optical thickness toward line and source
             center given in parentheses.
\end{table*}
\begin{table}
\caption{Models results for various CO abundance structures.}\label{sum3}
\begin{tabular}{lllll} \hline\hline
        & ISRF   & $T_{\rm de}$ & $I(^{13}$CO$)$$^{b}$ & $I($C$^{17}$O$)$$^{b}$ \\ 
        &        & [K]  & K~km~s$^{-1}$    & K~km~s$^{-1}$       \\ \hline
MMS6    & 1000   &$\ldots$& 53.6 (40)   & 11.1 (1.5)   \\
        & 1000$^{a}$   &  30~K  & 40.3 (31)   & 4.72 (1.2)   \\
        & 1000   &  35~K  & 26.2 (27)   & 2.63 (1.0)   \\
MMS9    & 1000   &$\ldots$& 64.3 (6.7)  & 7.13 (0.26)  \\
        & 1000$^{a}$   &  30~K  & 54.0 (5.6)  & 4.56 (0.21)  \\
        & 1000   &  35~K  & 30.2 (4.3)  & 1.88 (0.16)  \\
FIR4    & 1000   &$\ldots$& 81.3 (11)   & 16.4 (0.43)  \\
        & 1000   &  30~K  & 71.0 (9.3)  & 10.6 (0.35)  \\
        & 1000   &  35~K  & 48.1 (7.4)  & 5.68 (0.28)  \\ 
        & 5000$^{a}$   &  40~K  & 66.7 (6.0)  & 5.83 (0.23)  \\ 
        & 10000  &  45~K  & 67.0 (4.8)  & 4.92 (0.18)  \\ \hline
\end{tabular}

For all models, a depleted abundance $X_{\rm D}$ has been used
  which is a factor 100 lower than the canonical CO abundance
  $X_0=10^{-4}$. $^{a}$Best fit model. $^{b}$See footnote b, Table~2.
\end{table}

\begin{figure}
\resizebox{\hsize}{!}{\includegraphics{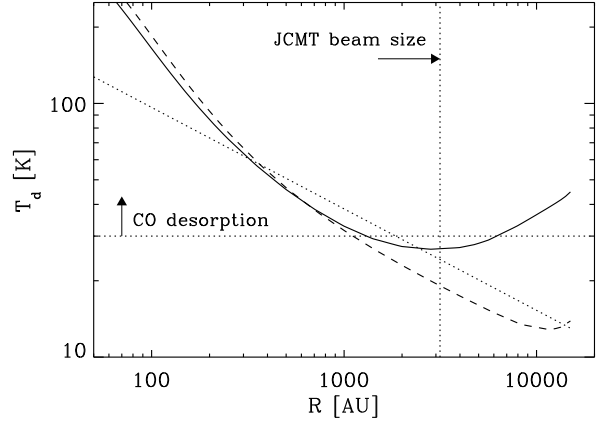}}
\caption{Temperature profiles for MMS6: models for the enhanced
($\times 1000$) and standard ISRF are indicated by the solid and
dashed lines, respectively. The dotted line indicates the temperature
profile for an optically thin envelope from Eq.~(\ref{e:chandlertemp})
corresponding to $q=0.4$ (or $\kappa \propto \nu^\beta$ with
$\beta=1$). The size of the JCMT beam is indicated by the vertical
dotted line, the CO evaporation temperature (see discussion in
Sect.~\ref{co_depletion}) with the horizontal dotted
line.}\label{f:mms6_temp}
\end{figure}

Fig.~\ref{f:mms6_temp} compares the dust temperature profile for MMS6
with the JCMT beam size and the results for an optically thin
envelope. The figure shows that a single pointing toward the source
center will barely resolve the region where the dust temperature
starts to increase due to the impact of the ISRF. This radius is
therefore also very well-matched to the JCMT beam, which will
selectively pick-up the outermost region. Since the envelope is
optically thick, the temperature will be higher in the innermost
region than predicted by the optically thin approximation, but
decrease much more rapidly outside a few hundred AU. The luminosity
derived from Eq.~\ref{e:lumargument} above is therefore a lower limit
to the required luminosity, further underscoring the need for a strong
interstellar radiation field. Finally, it can be seen that the
temperature does drop below $\approx 30$~K in the intermediate region
at radii $\approx$~1000--6000~AU, where CO depletion may
occur. However, in the outermost region where the temperature
increases again, CO will desorb rapidly from dust grains resulting in
``standard'' CO abundances.

\begin{figure}
\resizebox{\hsize}{!}{\includegraphics{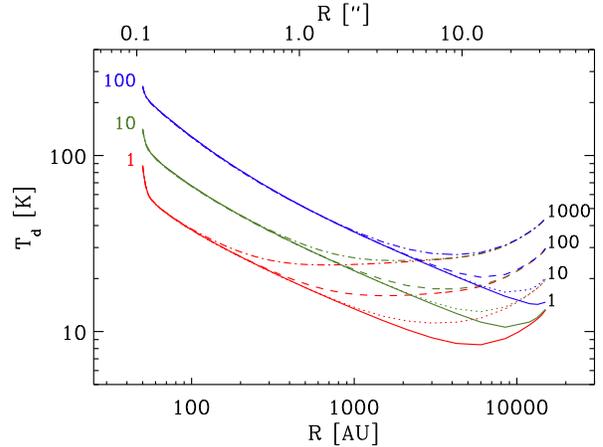}}
\caption{Dust temperature profiles as function of radius (lower
  X-axis) and density (upper X-axis) for a $n\propto
  r^{-1.5}$-envelope with inner and outer radii of 50 and 15,000~AU,
  respectively and $n_0$=2.7$\times 10^{8}$~cm$^{-3}$. Models are
  shown for luminosities of the inner source of 1, 10, and
  100~$L_\odot$ (red, green and blue lines) and $\chi_{\rm ISRF}$ of
  1, 10, 100 and 1000 (solid, dotted, dashed and dashed-dotted lines)
  as indicated in the left and right part of the figure,
  respectively.}\label{f:temp_inout}
\end{figure}

Fig.~\ref{f:temp_inout} compares the temperature profiles for varying
strengths of the interstellar radiation field and internal
luminosities. The figure clearly illustrates the competition between
the interstellar radiation field and the internal source luminosity in
defining the temperature structure - from the $L_{\rm int} = 100
L_\odot$ model with a standard interstellar radiation field where only
the outer region ($r > 10,000$~AU) of the envelope is affected by the
external heating to the $L_{\rm int} = 1 L_\odot$ model with
$\chi_{\rm ISRF}=1000$ where the interstellar radiation field
dominates the temperature in to radii of $\sim 100$~AU. For such a
strong ISRF even a core without a central source or heating (or of
very low luminosity) remains at high temperatures $\gtrsim 20$~K all
the way through to its center.  Currently, the fluxes of these sources
shortward of 100~$\mu$m are not known and model fits have only been
made to the long-wavelength tail of the SED \citep[see][and references
  therein for data]{johnstone03}. It is therefore not possible to
separate the internal luminosity of the cores from the contribution
from the external radiation field, and the temperature structure at
the smallest radii therefore remains poorly constrained. Sensitive
infrared observations, e.g., from the Spitzer Space Telescope, are
needed to separate the contributions from the internal luminosities
and the dust heated by the external radiation field to the total
radiation emitted by the cores - such as previously done for giant
molecular cloud cores near the galactic center based on ISO-LWS data
by \cite{lis01}.

Another problem is the determination of the outer radius of the
cores. These cores are not formed in isolation but located in the
Orion ridge, which is also picked up at larger scales of the
brightness profiles as seen in Fig.~\ref{f:dustprofile}. The ridge is
found to have a density profile $n \propto r^{-2}$ on scales up to
60--300$''$ \citep{johnstone99}. It would therefore seem more
realistic to adopt an attenuated profile at the outer edge of 15,000
AU for the ISRF such as also suggested by the modeling by
\cite{shirley05} and \cite{evans05b335}. For a density profile
$n\propto r^{-2}$ the column density from the outer edge, $r_y$, to
infinity is $N = n(r_{\rm y})\,r_{\rm y} \propto r_{\rm
  y}^{-2}\,r_{\rm y}$, i.e., declining as $r_{\rm y}^{-1}$. For MMS6,
for example, this implies a maximum extinction between the outer edge
at 15,000 AU and the outside radiation field of $A_V \approx 6$. If
the outer radius of the cores is increased the resulting line
intensities are unchanged, however (see Table~\ref{sum2}): for the
envelopes with density profile power-law slopes, $p$, of 1.5--2.0, the
main contribution to the beam averaged column density is located at
radii corresponding to the beam size or smaller. The $^{13}$CO line
becomes optically thick at roughly the same radius (and temperature)
in the envelope.

\subsection{Lines}
Adopting the physical structure of each envelope from the dust models,
the line emission is then modeled using the Monte Carlo radiative code
of \cite{schoeier02}. This code was benchmarked together with other
line radiative transfer codes by \cite{vanzadelhoff02}. The molecular
data were taken from the database of \cite{schoeier03radex}. For each
model a turbulent broadening of 0.7-1.0~km~s$^{-1}$ was adopted, reproducing
the observed line widths. For the optically thick $^{13}$CO transitions
inclusion of a more detailed velocity field may change the line
opacity. On the other hand, if these objects indeed are young and the
inside-out collapse model can be used to describe the velocity field
or if the infall can be described by a power-law velocity field, the
outermost regions should be close to static and the $^{13}$CO 3--2 lines
dominated by turbulent broadening \citep[see, e.g., discussion
in][]{paperii}.

\begin{figure}
\resizebox{\hsize}{!}{\includegraphics{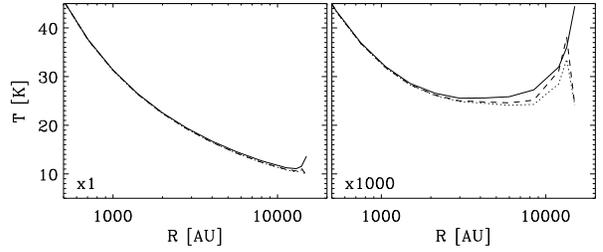}}
\caption{Comparison between dust and gas temperature for a standard
(left) and 1000$\times$ enhanced (right) interstellar radiation
field. In both panels the solid line is the dust temperature. The gas
temperature is indicated with the dashed line (assuming a turbulent
broadening 1.6~km~s$^{-1}$) and the dotted line (assuming a turbulent
broadening of 0.7~km~s$^{-1}$).}\label{temp_decoup}
\end{figure}

To first order the gas and dust temperatures have been assumed to
  be perfectly coupled.  This is justified by the high densities --
  generally $\gtrsim 10^5$~cm$^{-3}$ for radii $\lesssim 7500$~AU --
  inferred throughout most of the envelope, which leads to strong
  thermal coupling between the gas and dust via collisions
  \citep{doty97}. In a strong external ISRF the dust temperature may
  be less closely coupled to the gas temperature due to photoelectric
  heating and UV pumping of H$_2$.  We tested the effect of the ISRF
  by calculating the gas temperature as described by
  \cite{doty97}. The results are shown in Fig.~\ref{temp_decoup}. As
  can be seen, even in these strong external radiation fields, the
  densities studied here lead to dominant collisional coupling, and
  thus only marginal ($\lesssim$~a few K) differences between the dust
  and gas temperatures.

For the first radiative transfer iteration, the abundances are assumed
to be constant adopting a canonical CO abundance $X_0 = 1\times
10^{-4}$ and $^{12}$C:$^{13}$C and $^{16}$O:$^{17}$O ratios of 70 and
1950, respectively. The exact CO abundance will affect the derivation
of the strength of the ISRF somewhat as it will regulate the location
in the envelope where the observed $^{13}$CO emission becomes optically
thick. On the other hand, as the $^{13}$CO 3--2 emission \emph{does}
become optically thick the exact abundance structure in the interior
of the envelope where depletion may occur \emph{does not} affect the
constraints on the ISRF (see also discussion below).

\begin{figure}
\resizebox{\hsize}{!}{\includegraphics{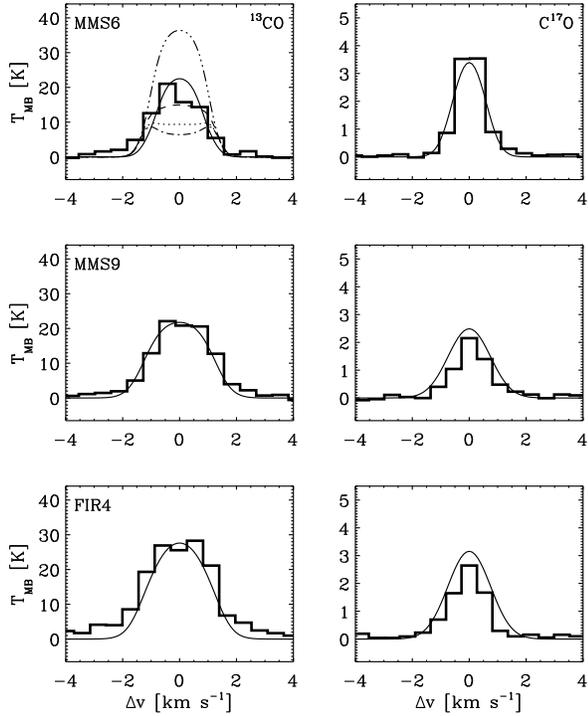}}
\caption{Modeled and observed line profiles for $^{13}$CO (left)
    and C$^{17}$O (right) $J=3-2$ for MMS6 (upper), MMS9 (middle) and
    FIR4 (lower). For MMS6 $^{13}$CO models are shown for 1, 10, 100,
    $10^3$ and $10^4$ times the standard ISRF with black
    dashed-dotted, dotted, dashed, solid and dashed-3$\times$dotted
    lines, respectively. For the remaining sources the best fit models
    from Table~\ref{sum3} and \ref{t:bestfit} are shown.}\label{f:mms6_13comodels}
\end{figure}

As shown in Fig.~\ref{f:mms6_13comodels}, an interstellar radiation
field 1000 times stronger than the standard field is needed to account
for the peak temperatures of the observed $^{13}$CO spectra and fill out
the self-absorption dip. It gives relatively Gaussian profiles and
line widths of 2-2.5~km~s$^{-1}$ (FWHM) in agreement with observations. It
is worth emphasizing here the differences between the observations of
the $^{13}$CO and C$^{17}$O lines: the C$^{17}$O lines are significantly narrower
(1.0-1.5~km~s$^{-1}$; FWHM) than the $^{13}$CO lines \citep{johnstone03} (see
Fig.~\ref{f:mms6_13comodels}). This simply reflects the fact that the
C$^{17}$O lines are optically thin and is confirmed by the
models. Table~\ref{sum2} and \ref{sum3} summarize the results of the
models for each source and Table~\ref{t:bestfit} lists the parameters
for the best fit model for each source.
\begin{table}
\caption{Best fit parameters.}\label{t:bestfit}
\begin{tabular}{llll}\hline\hline
Source                                   &  MMS6   &  MMS9   &  FIR4      \\ \hline
$\chi_{\rm ISRF}$                        & 1$\times 10^{3}$ & 1$\times 10^{3}$ & 1$\times 10^{4}$   \\
CO depletion, $X_0 / X_{\rm D}$          & 100     & 100     & 100         \\
Desorption temp., $T_{\rm ev}$           & 30~K    & 30~K    & 45~K        \\
$[$p-H$_2$CO$]$                          & 2.8$\times 10^{-10}$ & 3.6$\times 10^{-10}$ & 5.5$\times 10^{-10}$ \\ \hline
\end{tabular}
\end{table}
\section{Discussion}\label{s:isrf}\label{results}\label{discuss}
In the previous section, the contribution from the external heating of
the cores through the interstellar radiation field has been determined
using the established framework of dust and gas radiative
transfer. However, some simplifying assumptions were made and this
section discusses how these models can be refined by taking the
importance of chemistry into account.

\subsection{CO depletion and desorption temperature}\label{co_depletion}
\cite{jorgensen02} derived the CO abundances for a sample of $\sim$20
pre- and protostellar objects, among them NGC~1333-IRAS~2A. Clear
signs of CO freeze-out were observed, with the highest degree of
depletion found in the objects with the most massive envelopes. Models
of the C$^{17}$O lines suggest that freeze-out also occurs in the
intermediate mass envelopes studied in this paper: adopting a constant
canonical CO abundance of $10^{-4}$ as was done for $^{13}$CO in the first
iteration above, overproduces the observed C$^{17}$O 3--2 line intensities
by about a factor three. It therefore seems that, at least in the
region where the C$^{17}$O lines are sensitive, the abundances must be
lower. Since the C$^{17}$O 3--2 line generally is optically thin
throughout the studied envelopes, the line is sensitive to the
chemical structure in the interior of the envelope.

The freeze-out of CO -- and thus its abundance -- depends on
temperature and density in the envelope and is therefore not constant
but rather varies with radius. A CO desorption temperature higher than
20~K is suggested for low-mass protostellar cores from Monte Carlo
modeling of multi-transition data and direct imaging through
high-resolution interferometer observations. These observations show
that CO evaporates off the dust grains in the innermost envelopes
wherever the temperature becomes higher than 35--40~K
\citep{jorgensen02,l483art,coevollet}. Detailed chemical models of the
low-mass protostar IRAS~16293-2422 by \cite{doty04} suggest a
desorption temperature of CO of 20--40~K. 

To simulate this effect, we introduce a simple piece-wise constant
``drop'' abundance structure: we fix the abundances in three regions
of the envelope, adopting a standard [CO/H$_2$] abundance of
$X_0=1\times 10^{-4}$ in the inner and outermost region of the
envelope where CO desorbs due to the temperature increasing above the
CO desorption temperature $T \ge T_{\rm ev}$ and a depleted abundance
$X_D$ by two orders of magnitude in the region where the temperature
drops below $T_{\rm ev}$. This model then only has one free parameter,
$T_{\rm ev}$, which can be fit by comparison to the $^{13}$CO and
C$^{17}$O line intensities. In fact we are not sensitive to the
temperature, $T_{\rm ev} ({\rm in})$, at the innermost of these radii
as the region inside this contributes only a negligible amount of
material to the beam averaged C$^{17}$O column density. Similar
abundance profiles have been suggested for low-mass protostellar
envelopes: in these envelopes the densities in the exterior are low so
that the timescale for CO freeze-out is longer than the ``age'' of the
core, which results in un-depleted abundances there. This is less of
an issue for the Orion cores, however, since for these sources the
densities are high enough at the edge that CO depletion does occur in
the lifetime in the core, provided the temperature is low enough.

For MMS6 and MMS9 the $^{13}$CO and C$^{17}$O line intensities are well
reproduced by these structures where the abundance decreases when
(going from the inside) the temperatures decreases below 30~K at
$\approx$1300~AU (MMS6) and $\approx$2300~AU (MMS9) and the abundance
increases again when the temperature increases above 30~K (due to the
external heating) at radii of $\approx$6400~AU (MMS6) and
$\approx$5300~AU (MMS9) (Table~\ref{sum3}). $^{13}$CO still becomes
optically thick in the outermost envelope and is thus not sensitive to
the depletion region, whereas the C$^{17}$O lines require an abundance
decrease by 2 orders of magnitude. For both sources, models with
$T_{\rm ev}=30$~K are found to provide good fits to the data. A very
interesting point is illustrated by the fits to FIR4: for this source
it is not possible to fit the $^{13}$CO and C$^{17}$O simultaneously
with $\chi_{\rm ISRF}=1000$: for $T_{\rm ev}=30$~K the C$^{17}$O line
is overproduced by a factor 2 by the envelope models whereas the
$^{13}$CO 3--2 line is well-reproduced. Increasing the CO desorption
temperature to 35~K (i.e., increasing the size of the CO depletion
zone) lowers the C$^{17}$O 3--2 line intensity to the observed value,
but due to the $^{13}$CO 3--2 line being only moderately thick, the
temperature probed by this line simultaneously drops and the line
intensity correspondingly decreases by 35\%. If $\chi_{\rm ISRF}$ is
increased by a factor 10 the envelope naturally becomes warmer, never
dropping below 35~K. However, for such a model both $^{13}$CO and
C$^{17}$O line intensities are well-reproduced with a desorption
temperature of 45~K (corresponding to radii of 1700~AU and
7700~AU). This suggests a difference in the environment of FIR4 and
the two other cores, with a stronger contribution of the ISRF
impacting FIR4. Such a difference is in fact reasonable: FIR4 is
located at about half the distance from the illuminating
Trapezium stars compared to MMS6 (Table~1), so from pure
geometric dilution an enhancement of the ISRF by about a factor
  of 4 is to be expected. An even larger enhancement difference
between FIR4 and the two other sources is possible if the radiation
field is attenuated by any intervening cloud material.

\begin{figure}
\resizebox{\hsize}{!}{\includegraphics{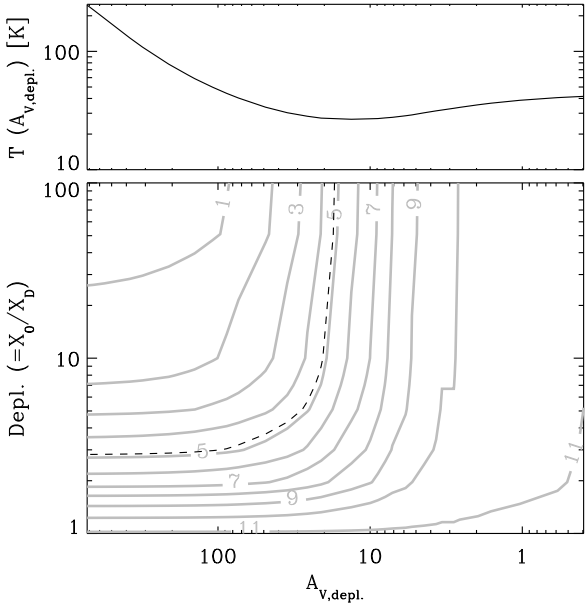}}
\caption{Modeled line intensities in ``jump models'' for MMS6 where
  the C$^{17}$O abundance is decreased to $X_{\rm D}$ from the outside
  inwards to a specific radius corresponding to an extinction $A_{\rm
    V,depl.}$ and a canonical abundance $X_0$ of 5.1$\times 10^{-8}$
  (corresponding to a CO abundance of 10$^{-4}$) within this
  radius. The density and temperature profiles are otherwise
    similar to the best fit model for MMS6. The contours are given in
  steps of 1~K~km~s$^{-1}$ ranging from 1 to 11~K~km~s$^{-1}$ as labeled. The
  dashed line indicates the observed line intensity for MMS6. The
  upper panel shows the temperature at the radius of the jump at
  $A_{\rm V,depl.}$.}\label{co_photodiss}
\end{figure}

Rather than the $^{13}$CO and C$^{17}$O lines probing material with different
levels of freeze-out, isotope selective photodissociation may result
in changed abundances for the two species. As argued by
\cite{vandishoeck88}, the photodissociation is dominated by line
absorption (rather than continuum absorption) so that self-shielding
and mutual shielding lines of more abundant isotopes and H$_2$ are
important for regulating the photodissociation rate of the various
isotopes. This can in principle affect the abundance ratios of the
isotopic species, e.g., by decreasing the less abundant species such
as C$^{17}$O relative to $^{13}$CO compared to the ``standard isotope
ratios''. Fig.~\ref{co_photodiss} shows the predicted C$^{17}$O 3--2
line intensities for models with a low C$^{17}$O abundance from the
outside of the envelope in to a radius corresponding to a specific
column density (or extinction, $A_{\rm V,depl.}$) in the envelope
simulating photodissociation of C$^{17}$O. Inside this radius the
C$^{17}$O abundance is increased to its canonical value assuming the
standard isotope ratios. As shown the C$^{17}$O abundance has to
either be lower by one to two orders of magnitude into an $A_{\rm
  V,depl}$ of about 15--20 or lower by a factor of about 3 into an
$A_{\rm V,depl.}\gtrsim 50$. Photodissocation will most likely only
affect the outermost ($A_V \lesssim$~a~few) envelope and only decrease
the abundance by a few, it is therefore not sufficient to invoke
isotope selective photodissociation to explain the differences between
the material probed by $^{13}$CO and C$^{17}$O.

\subsection{H$_2$CO emission}
An interesting test of the derived models is how they account for the
emission from other molecular species. In particular, ratios between a
few key lines of H$_2$CO have been suggested as a useful temperature
probe \citep{mangum93} and may therefore be particularly sensitive to
the impact of the ISRF. \cite{johnstone03} assumed constant densities
and temperatures for each of the protostars in this paper and found
temperatures of $<$50~K for MMS6, 30~K for MMS9 and 80-90~K for
FIR4. For comparison the mass-weighted temperatures,
\begin{equation}
\langle T
\rangle=\int T(r)\,n(r)\,4\pi r^2\, {\rm d}r/\int n(r)\,4\pi r^2\,
{\rm d}r,
\end{equation}
from the derived envelope models are 36~K for MMS6 and MMS9 and 51~K
for FIR4 (see Table~\ref{sum2}). The variations of these temperatures
between the studied sources therefore do appear to reflect the
differences in the external heating, with FIR4 again being
significantly warmer than the remaining sources.

The enhanced ISRF and thus more strongly varying temperature profile
may, however, produce a discrepancy for sensitive temperature probes
such as the ratios of the H$_2$CO lines. To quantify this issue,
simple constant abundance models were constrained by comparison to the
observed H$_2$CO line intensities. Table~\ref{h2co_modeltab} lists the
best fit models for models with both $\chi_{\rm ISRF}=1$ and the best
fit, enhanced ISRF for each source. It is clear that the models with
the enhanced ISRF provide significantly better fits to the observed
H$_2$CO line intensities. The abundances are found to be remarkably
similar at 3--5$\times 10^{-10}$. For comparison \cite{hotcoresample}
found constant abundances of 3$\times 10^{-11}$--4$\times 10^{-9}$ for
a sample of 18 low-mass protostars.

The Orion cores are also interesting in the context of their H$_2$CO
abundance structures: enhanced H$_2$CO abundances have been suggested
for low-mass protostars in regions where ices evaporate from the dust
grains at temperatures higher than 90--100~K although the evidence
still is inconclusive (see discussions in \cite{maret04} and
\cite{hotcoresample}). Since the cores in Orion have higher
luminosities, the innermost hot ($T > 90$~K) region is expected to be
larger, and, despite their greater distances, these hot cores would be
less diluted in the single-dish beam than the sources studied by
\cite{maret04} and \cite{hotcoresample} (diameters of $1.0-1.5''$ for
the Orion sources compared to the $\lesssim 0.5''$ for the low-mass
protostars). We find, however, that the observed H$_2$CO lines are
best fit by a constant abundance or a modest abundance jump at
best. This conclusion is only based on a few lines, however, some of
which may become optically thick and therefore more sensitive to the
enhanced temperatures due to the external radiation field.  Also,
chemistry in the outer envelope, especially freeze-out, may affect the
line intensities measured for H$_2$CO \citep{hotcorepaper}. Finally we
reemphasize that it is not required that these sources have a luminous
internal source from either the dust modeling or the gas modeling,
which also is important in this aspect. Further studies of high
excitation transitions at high angular resolution are needed to fully
address these issues.

\begin{figure}
\resizebox{\hsize}{!}{\includegraphics{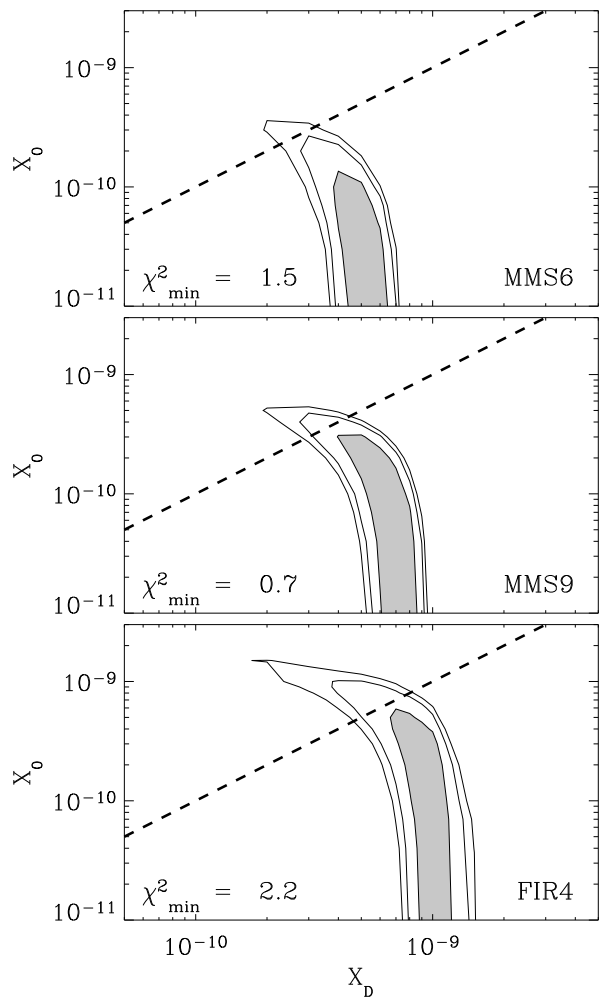}}
\caption{$\chi^2$-confidence plots for fits to the H$_2$CO lines
  toward MMS6, MMS9 and FIR4. The 68.3\% ('1$\sigma$'; also
  grey-scale), 90\% and 95.4\% ('2$\sigma$') levels are shown. The
  abundance structures are assumed to follow those of the best fit CO
  models, i.e., with a change in abundance from the canonical value,
  $X_0$ to $X_{\rm D}$ in the intermediate regions where the
  temperatures drop below 30~K (MMS6 and MMS9) and 45~K (FIR4). The
  dashed lines indicate constant abundance models. The minimum
  $\chi^2_{\rm red}$ for the models for each source is given in the
  lower left corner of each panel. Note that in all cases the
    best fit model has a higher $X_{\rm D}$ than $X_0$, but also that
    constant abundances are consistent with the data on the 2$\sigma$
    level.}\label{h2co_chi}
\end{figure}

The large spread in H$_2$CO abundances for the low-mass protostellar
cores \citep{hotcoresample} may be related to CO depletion:
\cite{maret04} and \cite{hotcoresample} find that the average
abundances of CO and H$_2$CO are correlated in the outer cold
envelopes of low-mass protostars and high resolution interferometer
observations indicate that H$_2$CO follows a similar abundance
structure as CO \citep{hotcorepaper}. H$_2$CO abundances for the three
Orion sources were therefore also calculated with drop abundance
structures based on the CO results, i.e., varying the undepleted and
depleted H$_2$CO abundances ($X_0$ and $X_{\rm D}$, respectively) and
adopting $T_{\rm ev}$ found for the fits to the CO
lines. Fig.~\ref{h2co_chi} shows the $\chi^2$ confidence
plots. Interestingly a drop in abundance is not indicated by the
observed line intensities. Constant abundance or small increases in
abundance within the CO depletion region are allowed by the 2$\sigma$
confidence levels for the observed sources.
\begin{table}
\caption{Summary of best-fit constant abundance H$_2$CO models with
  differing contributions by the interstellar radiation field. All
  considered transitions are from para-H$_2$CO.}\label{h2co_modeltab}
\begin{center}\begin{tabular}{llll} \hline\hline 
Source   & ISRF & Abundance (para-H$_2$CO)                & $\chi^2_{\rm red}$ \\ \hline
MMS6     &  1   & 2.3$\times 10^{-10}$                            & 5.2 \\
         &$10^3$& 2.8$\times 10^{-10}$                            & 2.9 \\
MMS9     &  1   & 2.2$\times 10^{-10}$                            & 5.1 \\
         &$10^3$& 3.6$\times 10^{-10}$                            & 1.8 \\
FIR4     &  1   & 2.2$\times 10^{-10}$                            & 6.7 \\
         &$10^4$& 5.5$\times 10^{-10}$                            & 3.0 \\
NGC~1333-IRAS~2A &  0   & 3.9$\times 10^{-10}$                            & 1.3$^{b}$ \\ \hline
\end{tabular}\end{center}

$^a$2$\sigma$ upper limit. $^{b}$For four fitted lines, including the
H$_2$CO $5_{24}-4_{23}$ transition at 363.946~GHz
\citep{hotcoresample}.
\end{table}

\section{Evolutionary implications}\label{s:implic}
The points discussed above have interesting implications related to
the evolution of the cores. The first concerns their history: as shown
in Fig.~\ref{f:temp_inout} even for a pre-stellar core, i.e., a core
with no internal source of heating, the temperature never decreases
below 25~K in such an enhanced external radiation field considered in
this paper. Under these conditions a given CO molecule would desorb
again almost instantaneously after freeze-out so that only little
depletion should be expected. It is, however, clear that CO \emph{is}
frozen-out in parts of the observed envelopes, suggesting that these
regions must have been at lower temperatures earlier in the evolution
of the cores.

An explanation may be that the strength of the UV field in the Orion
region has changed significantly during the evolution of the studied
cores.  Plenty of circumstantial evidence suggests that the Trapezium
cluster, which produces the majority of the local UV photons, is
relatively young \citep{odell01}.  For example, \cite{palla01}
determine the age for $\Theta^1\,$Ori\,B to be less than 100,000
years. The most massive star in the Trapezium, $\Theta^1\,$Ori\,C,
shows considerable variability in its X-ray emission \citep{gagne97} a
likely indicator of extreme youth.  Most, if not all of the low-mass
stars in the Trapezium cluster are still proplyds, despite losing mass
from their disks due to UV induced photoevaporation at rates of
$10^{-7}$ to $10^{-6}\ M_\odot\,$yr$^{-1}$
\citep{bally98,johnstone98}. Assuming that each disk has a minimum
Solar nebula mass of $\approx 10^{-2}\ M_\odot$, this implies a
lifetime of $<10^5\,$yr since the initiation of evaporation. Finally,
the H$\alpha$+\ion{O}{[III]} arcs observed around young low-mass stars
near the high mass Trapezium stars trace out a wind driven bubble the
size of which is very small, again suggesting extreme youth
\citep{odell97,bally98}.

\begin{figure}
\resizebox{\hsize}{!}{\includegraphics{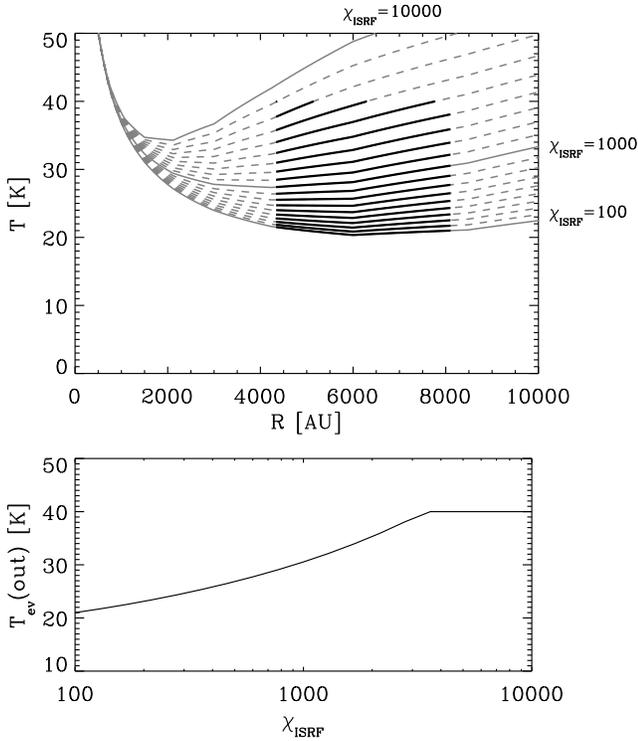}}
\caption{The evolution of the temperature profiles (grey lines) for a
  100~$L_\odot$ core with density profile as in
  Fig.~\ref{f:temp_inout}. Each line corresponds to increasing
  $\chi_{\rm ISRF}$ ranging from 100 to 10$^4$ in logarithmic
  steps. The grey solid lines indicate the temperature profiles for
  $\chi_{\rm ISRF}=100$, 1000 and 10,000. The thick black line in each
  plot indicates the radii where CO was frozen out for $\chi_{\rm
    ISRF}=100$. In the lower panel the evolution of the inferred
    $T_{\rm ev}({\rm out})$ with increasing ISRF strength is
    shown.}\label{evol_isrf}
\end{figure}

If the CO molecule is frozen-out at lower temperatures it may stay
bound to the dust grains at the higher temperatures observed at
present day, such as seen in low-mass protostellar envelopes where CO
is expected to freeze-out in the cold ($T< 15$~K) stages and evaporate
as the protostar starts heating the envelope, increasing the
temperature above $\approx 35-40$~K. One explanation for this
according to laboratory experiments is that CO migrates into a porous
ice when heated \citep{collings03}. It is therefore also not clear
whether the determined values for $T_{\rm ev}$ in this paper
correspond to the radii where the ice was formed at a lower
temperature in an earlier stage or rather the radius where the
temperature is high enough so that it evaporates at present. In fact
the difference between FIR4 on one hand and MMS6 and MMS9 on the other
may reflect such effects: Fig.~\ref{evol_isrf} shows the evolution of
the temperature distribution in a 100~$L_\odot$ envelope with a similar
density profile as in Fig.~\ref{f:temp_inout} but for varying ISRF
strengths. At the earliest stages CO freezes out between radii where
the temperature is lower than 21~K, but as the ISRF strength increases
the same radii will correspond to increasing temperatures. As the
temperature then increases above $T_{\rm ev}$, CO desorbs and the
radius for the outer edge of the depletion zone moves inwards.

If this interpretation is true -- the cores had CO freeze-out in
  the outer regions when formed, and then had CO evaporate when the
OB stars ``turned on'' -- this will also have important chemical
implications in relation to the grain chemistry and for gas-phase
species regulated by the amount of CO. The H$_2$CO and CO correlation
can easily can be understood since the amount of CO and related C$^+$
and O is the main bottleneck for gas-phase H$_2$CO production
\citep{hotcorepaper}. In a gas-phase chemical model using the approach
of \cite{doty04}, and adopting the physical conditions
(incl. temperature and density) in MMS6, the H$_2$CO abundances
follows the CO abundance structure after $\sim 10^5$~years.

Although the abundance structures are similar in the Orion cores and
in the low-mass protostellar envelopes, the grain chemistry might be
significantly different. In both cases high abundances of, e.g., CO
are expected in the outer envelopes, but whereas this is caused by low
densities (and thus long timescales for freeze-out) in the low-mass
protostellar envelopes, it is a result of direct desorption (high
temperatures) in the Orion protostars. This has the consequence that
the timescale where CO is available for surface chemical reactions
e.g., hydrogenation forming CH$_3$OH and the rates of these reactions
are different in the Orion cores due to their dependence on
temperature.

It will naturally be very interesting to test the importance of any
photochemistry. In particular an increased CO abundance provides high
amount of atomic carbon, which will lead to higher abundances of for
example CN and CS in the outer envelopes relative to related species
such as HCN and SO.

\section{Conclusions}\label{s:conclusion}
We have presented an analysis of submillimeter continuum and line
observations toward three intermediate mass protostellar cores in the
Orion molecular cloud using detailed radiative transfer models. The
main conclusions of this study are:

\begin{itemize}
\item From simple luminosity and temperature considerations these
  sources must be subject to a strong external heating to reproduce
  observed $^{13}$CO 3--2 line temperatures and profiles.
\item Detailed radiative transfer modeling confirms this, constraining
  the ISRF enhancement to $10^3-10^4$ times the standard ISRF from the
  studied sources. Different strengths of the ISRF between
    MMS6 and MMS9 in one case and FIR4 in the other can be explained
    in part due to the latter being closer to the Trapezium stars
    supplying the strong UV field.
\item The C$^{17}$O 3--2 lines are optically thin and thus not
  sensitive to the same temperature enhancement. Differences
    between the $^{13}$CO and C$^{17}$O line widths reflect this difference
    between the optically thick and thin lines, which is well
    reproduced by the models. To reproduce the C$^{17}$O line intensities,
    significant CO depletion must have occurred in the part of the
    envelope where the temperature is lower than 30-45~K. A difference
    in the CO desorption temperatures derived for the sources,
    however, suggest an evolutionary difference with the OB stars (and
    thus UV field) evolving over similar timescales as the
    protostellar cores themselves.
\item Multi-transition H$_2$CO observations indicate high temperatures
in the cores and their line ratios can only be reproduced in models
with a strong ISRF. Constant abundances of 3$\times 10^{-10}$--5$\times 10^{-10}$
provide a good fit to the line intensities.
\end{itemize}

This work illustrates the necessity of establishing the environmental
impact for the evolution of pre- and protostellar cores - for example
in regions such as Orion where the impact of the radiation stars have
a large impact on newly formed low- and intermediate mass
protostars. As the above discussions also illustrate, detailed
radiative transfer models can be used to address some of these issues,
in particular if the chemistry is taken into account.

Additional mid-infrared observations of the cores can further
constrain the relative contributions of the internal and external
radiation field to the observed spectral energy distributions as it
has previously been done for cores near the galactic center
\citep{lis01}. Further continuum observations with (sub)millimeter
interferometers such as the SMA, CARMA and eventually ALMA could
further confirm the temperature structure of the envelopes at smaller
scales and in particular whether desorption of molecules are seen
toward their centers due to the heating by the central
protostar. Likewise in an extreme interstellar radiation field such as
that of the Orion further studies of the chemistry in the outer
envelopes and intervening material in the ridge are interesting to
fully understand the impact of the ionizing sources.  Understanding
these processes are important as the thermal balance and pressure of
the cores controls the conditions and outcome of the collapse and thus
establishing the properties of the newly formed stars.

\begin{acknowledgement}
We thank the referee, Neal Evans, for insightful comments which
greatly improved the paper. The research of JKJ was funded through a
Ph.D. stipend from the Netherlands Research School for Astronomy
(NOVA) and NASA Origins Grant NAG5-13050. This research was also
supported by a grant from the Natural Science and Engineering Research
Council of Canada to DJ and a grant for from The Research Corporation
(SDD). Astrochemistry research in Leiden is supported by an
NWO/Spinoza grant. DJ wishes to thank the Sterrewacht Leiden for its
kind hospitality during the past four summers, during which much of
this research was conducted.
\end{acknowledgement}

\end{document}